# CLASSIFFICATION OF MILD COGNITIVE IMPAIRMENT BASED ON DYNAMIC FUNCTIONAL CONNECTIVITY USING SPATIO-TEMPORAL TRANSFORMER

*Jing Zhang[1], Yanjun Lyu[1], Xiaowei Yu[1], Lu Zhang[2], Chao Cao[1], Tong Chen[1], Minheng Chen[1], Yan Zhuang[1], Tianming Liu[3], Dajiang Zhu[1]*

[1]Computer Science and Engineering, The University of Texas at Arlington, Arlington, TX, USA
[2]Department of Computer Science, Indiana University Indianapolis, IN, USA
[3]School of Computing, The University of Georgia, Athens, GA, USA

## ABSTRACT

Dynamic functional connectivity (dFC) using resting-state functional magnetic resonance imaging (rs-fMRI) is an advanced technique for capturing the dynamic changes of neural activities, and can be very useful in the studies of brain diseases such as Alzheimer's disease (AD). Yet, existing studies have not fully leveraged the sequential information embedded within dFC that can potentially provide valuable information when identifying brain conditions. In this paper, we propose a novel framework that jointly learns the embedding of both spatial and temporal information within dFC based on the transformer architecture. Specifically, we first construct dFC networks from rs-fMRI data through a sliding window strategy. Then, we simultaneously employ a temporal block and a spatial block to capture higher-order representations of dynamic spatio-temporal dependencies, via mapping them into an efficient fused feature representation. To further enhance the robustness of this feature representations by reducing the dependency of labeled data, we also introduce a contrastive learning strategy to manipulate different brain states. Experimental results on 345 subjects with 570 scans from the Alzheimer's Disease Neuroimaging Initiative (ADNI) demonstrate the superiority of our proposed method for MCI (Mild Cognitive Impairment, the prodromal stage of AD) prediction, highlighting its potential for early identification of AD.

## 1. INTRODUCTION

Alzheimer's disease (AD) is the most prevalent chronic neurodegenerative disorder, accounting for 60-70% of dementias, and 1/3 of elder Americans die with AD or other types of dementias. Clinically, AD manifests as a progressive loss of memory, insight, judgment, and language abilities [1]. But the pathological changes in brain may begin to develop decades before clinical symptoms become evident, therefore, early detection of AD at its prodromal stage – MCI is critical and thus attracting extensive efforts in clinical studies [2].

rs-fMRI technology uses blood-oxygen-level dependent (BLOD) signals to non-invasively assess intrinsic functional connectivity within the brain [3]. It is regarded as a promising biomarker for AD since functional brain changes are thought to occur before structural alterations [4]. Functional connectivity (FC), defined as the interdependence of neuronal activation patterns between the BOLD signals of two brain regions [5], has been studied as potential biomarkers of brain disorders, such as AD and MCI. Most FC analytic approaches treat FC as static, assuming remains the connectivity does not change throughout the entire scanning period, known as static FC [6]. However, this assumption overlooks the brain's temporal dynamics, failing to capture the dynamic interactions in brain activities. Numerous studies have demonstrated that even during resting states, FC exhibits significant temporal variations [7], which are crucial for a more comprehensive understanding of brain pathology. Therefore, recent studies have increasingly focused on dFC to better characterize these temporal changes between brain regions, exploring their associations with brain diseases [8, 9, 10]. For example, support vector machine (SVM), Principal Component Analysis (PCA), and Independent Component Analysis (ICA) have been used to find temporal changes with fMRI. However, they are limited to characterizing dynamic patterns effectively due to their linearity and/or shallow nature. In recent years, deep learning (DL) models have emerged as powerful tools for a wide range of predictive tasks, demonstrating the ability to capture higher-level nonlinearities and learn representations that facilitate model training with minimal or no need for feature selection [11]. DL models also showed promising performance in diverse fMRI applications [12]. Transformer, originally developed for natural language processing (NLP), has become a milestone in artificial intelligence due to its proficiency in capturing temporal sequences and is widely applied to disease diagnosis [13, 14]. However, there is a lack of research on applying the transformer to understand how functional brain dynamics relate to our cognitive abilities and are affected by brain disorders.

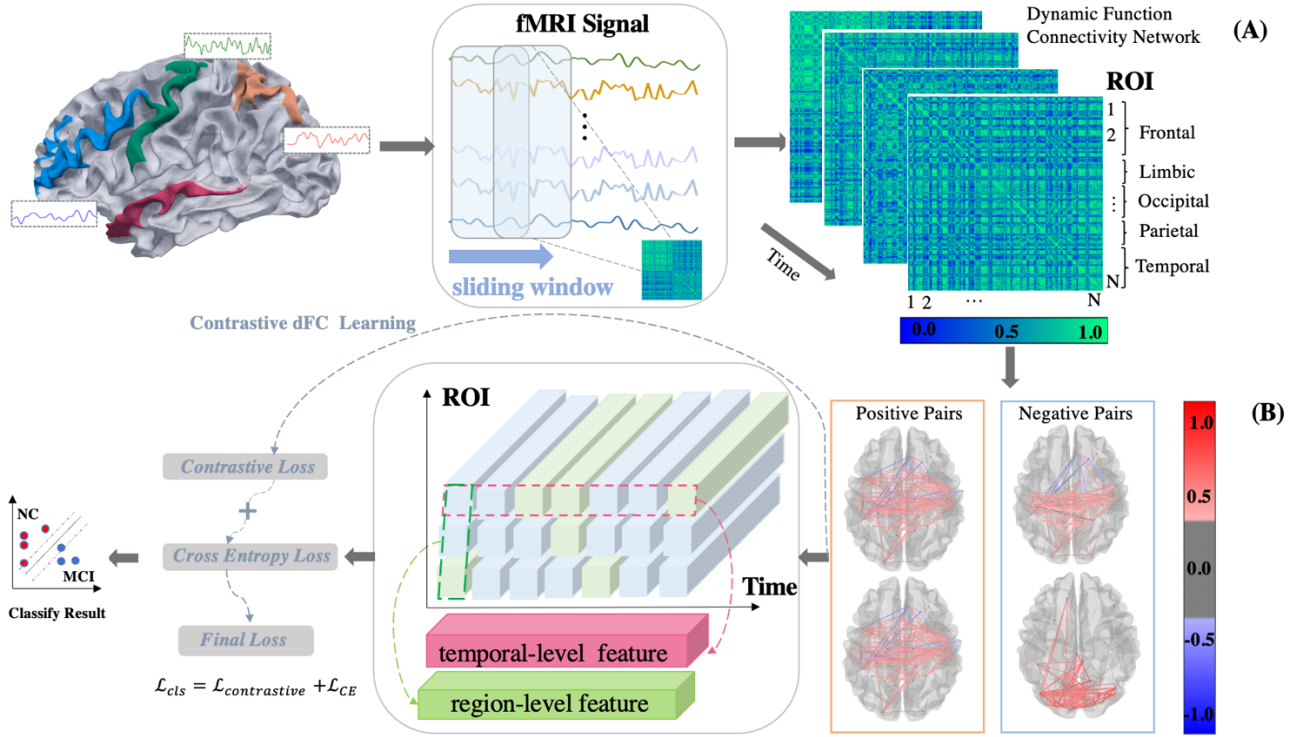

**Fig. 1.** Illustration of proposed Spatial-Temporal Transformer based on Contrastive Learning framework for brain disease classification with dFC, consisting of **(A)** imaging preprocessing and dynamic characterization of whole-brain connectivity construction. **(B)** Contrastive dFC pairs construction and feature extraction at Temporal and Region levels via Spatial-Temporal block. Both cross-entropy loss and contrastive loss are employed to optimize jointly.

To bridge this gap, in this work, we propose a novel learning framework for brain disease research shown in Fig. 1. Specifically, we first use the sliding window strategy to construct dFC networks and further use a spatial-temporal block based on transformer architecture to jointly learn higher-order representations of both region-level and temporal-level features. Furthermore, we generate contrastive dFC pairs based on diagnosis and introduce a contrastive learning [15] method to fully leverage the inherent information within the data itself, addressing a common data scarcity challenge in the medical domain. Compared with existing methods, our proposed framework shows promising classification performance compared to recent studies.

## 2. METHOD

### 2.1. Subjects and Image Preprocessing

We used 345 subjects of the ADNI dataset [16], which included both the normal control (NC) group (88 males, 137 females; 73.21 ± 7.66 years) and 120 subjects from the MCI group (70 males, 50 females; 74.64 ± 8.32 years). Each subject's resting-state fMRI (rs-fMRI) underwent the same standard pre-processing procedures as detailed in [17, 18]. In brief, we employed the FMRIB Software Library (FSL) FEAT to perform spatial smoothing, slice time correction, temporal pre-whitening, global drift removal, and band-pass filtering (0.01-0.1 Hz). The brain's cortical regions were identified using the Destrieux Atlas [19], and subsequently, the brain space of each subject's fMRI scan was segmented into 148 regions of interest (ROIs) after the exclusion of two unknown and two empty areas. For each subject, the average rs-fMRI time series from the BOLD signals across all ROIs were computed.

Next, we constructed the dFC network using an overlapping sliding window strategy, As shown in Fig. 1A. For each subject with $N$ ROIs, we divide the time series into $T$ overlapped windows. Within each $t$-th time window $(t = 1, \cdots, T)$, we constructed a functional connectivity matrix, $M_T$, representing the Pearson's correlation coefficient (PCC) of BLOD between any two specific ROIs. The length of each window was set to $L$ and the sliding step size to $S$. In constructing the dynamic FC networks, we empirically set the fixed length of the time window to $L = 70$ and the sliding step size to $S = 2$, which configuration has been proved to effectively capture dynamics [20]. This computation was repeated iteratively, shifting the window by the designated step each time, generating a connectivity time course. In this way, we finally compile a set of $T$ dFC networks, represented as $M = M_1, M_1, \cdots M_T$, which captures the dynamic characterization of whole-brain connectivity for each subject over time.

## 2.2. Model

As shown in Fig. 1B, our proposed framework consists of two key components: the first focuses on learning higher-order representations of dynamic Spatio-Temporal dependencies by incorporating a temporal block and a spatial block (detailed in Fig. 2). The second aims at generating contrastive dFC pairs and leveraging the inherent information within the data itself (positive and negative dFC pairs) to generate supervisory signals using a contrastive learning strategy.

*2.1.1. Temporal-level and Region-level Feature Extraction*

The temporal block is primarily responsible for capturing the dynamic changes over time across the various ROIs, simultaneously, the spatial block focuses on the spatial relationships among the ROIs. Specifically, each block consists of a transformer layer and a convolutional layer. In our work, the matrices $M_i$ can be represented as spatial-level feature matrices $M_i^s \in R^{m \times n}$ and temporal-level feature matrices $M_i^t \in R^{n \times m}$, where $M_i^s = (M_i^t)^\top$. Here, $m$ and $n$ denote the ROIs and timestamps, respectively. The input matrix $M$ is linearly projected to queries $Q$, keys $K$, and values $V$, and the transformer learning process can be formulated as:

$$Q = MW^Q, \; K = MW^K, \; V = MW^V \qquad (1)$$

$$MultiHead(Q,K,V) = Concat(head_1, \ldots, head_h)W^O \quad (2)$$

$$head_i(Q,K,V) = softmax\left(\frac{QK^\top}{\sqrt{d_k}}\right)V \qquad (3)$$

Here, $W^Q, W^K, W^V$ are learnable parameter matrices.

The Feedforward Neural Network (FNN) then applies two linear transformations with a ReLU activation function to the output of the multi-headed self-attention as:

$$FFN(x) = Relu(xW_1 + b_1)W_2 + b_2 \qquad (4)$$

where $x$ denotes the output of the previous layer, and $W_1, W_2, b_1, b_2$ are trainable parameters. Additionally, the residual connections with layer normalization are applied.

Next, the outputs from the temporal transformer and the spatial transformer are forwarded to their respective convolutional layers for further enhancing local connectivity and reducing dimensionality. Both temporal and spatial features are then concatenated and subjected to a global attention mechanism, generating the whole brain dFC features for the subsequent brain disease classification. The details are shown in Fig. 2.

*2.1.2. Contrastive dFC Learning*

Once the dFC networks for each subject are generated, we contrastive a set of pairs $\{(M_i, M_j)\}_{i,j=1}^p$ for p subjects,

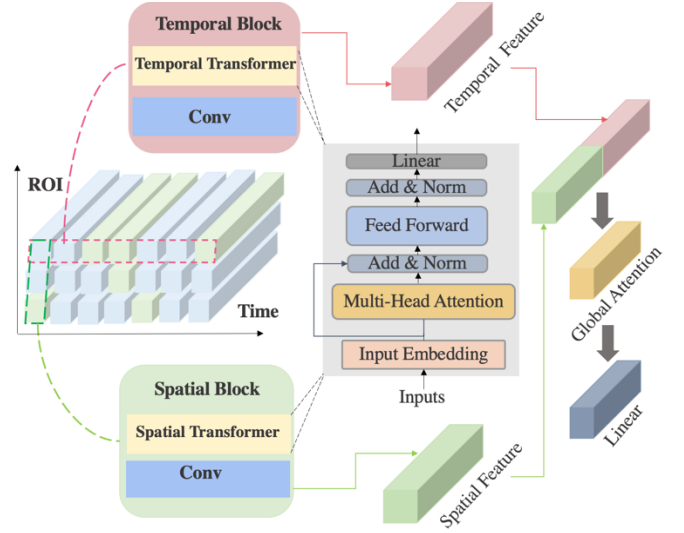

**Fig.2.** Illustration of Temporal-Spatial block. Each block consists of a transformer layer and a convolutional layer. The temporal block captures dynamic changes over time across different ROIs as temporal-level features, while the spatial block focuses on the spatial relationships among ROIs as region-level features. Temporal-level and region-level features are then concatenated and passed through a global attention and linear layer to form a fused whole-brain representation.

where dFC generated from subjects with the same diagnosis status (e.g., both NC) form "same" pairs, designed to "attract" each other. Conversely, dFC networks from different subjects with different diagnoses (e.g., NC and MCI) form "different" pairs and "repel" each other. Consequently, the model transforms dFC networks into high-order representations. The goal of the contrastive learning process is to maximize the similarity of positive dFC pairs while minimizing the similarity of negative dFC pairs, effectively learning robust feature representations capable of distinguishing between different brain states. To achieve this, we define the cosine similarity between two dFC temporal-level and region-level vectors $V_i$ and $V_j$ as:

$$d(V_i, V_j) = \frac{V_i^T V_j}{\|V_i\|\|V_j\|} \qquad (5)$$

Then, the loss function for a positive pair of examples $(V_i, V_i^+)$ is as follows:

$$\mathcal{L}_{contrastive}(V_i, V_i^+) = -\log \frac{\exp(d(V_i, V_i^+)/\tau)}{\sum_{k=1}^{2N} 1_{[k \neq i]} \exp(d(V_i, V_i^+)/\tau)} \qquad (6)$$

Where $1_{[k \neq i]}$ is a binary indicator function indicating whether the pair is positive or negative (0 or 1), and $\tau$ is a temperature parameter. This loss is computed across all pairs, leveraging the inherent information within the data itself to generate supervisory signals without requiring additional manual label information. Thus, it maximally the use of available data, effectively addressing a common data scarcity challenge in the medical domain.

*2.1.3. identification*

In the classification stage, we employ a single-layer perceptron with a nonlinear function to map the low-dimensional embedding into the classification space. To optimize the parameters, we utilize the cross-entropy loss defined as follows:

$$\mathcal{L}_{CE} = -\frac{1}{N}\sum_{i=0}^{N} y_i \cdot log(\hat{y}_i) + (1-y_i) \cdot log(1-\hat{y}_i) \quad (7)$$

Where $y_i$ denotes the category label. To enhance our training process, we incorporated the contrastive loss $\mathcal{L}_{contrastive}$ defined in equations (5) and (6) to jointly optimize, resulting in the overall loss function:

$$\mathcal{L}_{cls} = \alpha\, \mathcal{L}_{contrastive} + \beta\, \mathcal{L}_{CE} \quad (8)$$

where $\alpha$ and $\beta$ are learnable hyperparameters.

## 3. EXPERIMENT

### 3.1. Experimental Setting

The training process involves 64 epochs with a batch size of 8, utilizing a single NVIDIA TITAN GPU. The initial learning rate was set at $2 \times 10^{-6}$, and a weight decay parameter of 0.2 was applied to mitigate the risk of overfitting. The AdamW optimizer was employed throughout the training. We adopted a subject-level 5-fold cross-validation strategy to ensure that scans from the same subject did not appear in both the training and testing sets. The performance of the framework was evaluated using three objective metrics, including accuracy (ACC), sensitivity (SEN), specificity (SPE), area under the ROC curve (AUC), and the F1 score.

### 3.2. Classification Results

To verify our proposed model, we compare it with other widely used methods in the literature listed in Table 1. Our results outperform most other studies. Furthermore, the proposed method incorporating contrastive learning strategy performs significantly better than the method without contrastive learning strategy on this data, showing an improvement of 9.7% in accuracy and 7.3% in F1 score when contrastive learning is applied.

### 3.3 Effectiveness of temporal-spatial information

To evaluate the effect of the temporal block and spatial block in our proposed framework, ablation experiments are performed: 1) Using the original static FC matrix directly to construct the input, followed by classification using convolutional layers (os-FC); 2) Employing only the spatial block to learn the spatial relationships among the ROIs for classification (S-only); 3) Using only the temporal block to learn the dynamic sequential information across the various ROIs for classification (T-only). The results are shown in Table 2. It can be seen that the effect of the model is the best only when both the temporal block and spatial block are used at the same time.

**Table 2.** Ablation study results

| Method | ACC | SEN | SPE | AUC | F1 |
|---|---|---|---|---|---|
| os-FC | 66.3 | 68.2 | 60.0 | 59.1 | 61.2 |
| S-only | 76.6 | 78.7 | 75.4 | 77.2 | 73.1 |
| T-only | 70.7 | 73.2 | 72.0 | 72.6 | 78.6 |
| **Ours** | **89.1** | **91.4** | **87.2** | **89.3** | **90.3** |

os-FC: original static FC;   S/T-only: Using only the spatial/ temporal block

**Table 1.** Performance comparison with other methods.

| Method | Sample size | Modality | Atlas | ACC | SEN | SPE | AUC | F1 |
|---|---|---|---|---|---|---|---|---|
| Yu et. al [21] | 50MCI, 49NC | Static FC | AAL | 84.4 | 91.2 | 78.5 | N/A | N/A |
| Feng et al. [22] | 56EMCI, 50NC | Dynamic FC | AAL | 81.1 | 85.7 | 76.0 | 79.3 | N/A |
| Lin et al. [8] | 50EMCI, 48NC | Dynamic FC | AAL | 84.5 | 84.8 | 80.4 | N/A | N/A |
| Wee et al. [23] | 29EMCI, 30NC | Dynamic FC | AAL | 79.7 | 75.9 | 83.3 | 79.2 | 78.6 |
| Li et. al [24] | 36MCI, 37NC | Static FC | AAL | 65.8 | 69.4 | 62.2 | 54.3 | N/A |
| Li et. al [24] | 36MCI, 37NC | Dynamic FC | AAL | 76.7 | 77.8 | 75.7 | 67.5 | N/A |
| Du et. al [25] | 95MCI, 48NC | Static FC | AAL | 80.1 | 78.4 | 81.8 | 87.7 | N/A |
| **Ours** | 120MCI,225NC | Dynamic FC | Destrieux | 82.4 | 83.1 | 79.8 | 82.0 | 83.0 |
| **Ours (cl)** | 120MCI,225NC | Dynamic FC | Destrieux | **89.1** | **91.4** | **87.2** | **89.3** | **90.3** |

cl: contrastive learning

## 4. CONCLUSION

In this paper, we introduce a novel framework that simultaneously learns the embedding of both spatial and temporal information within dFC using transformer architecture. Additionally, we introduce a contrastive learning strategy that not only reduces dependence on labeled data but also achieves improved classification results for identifying MCI from NC. Finally, we conduct ablation experiments to verify the effectiveness. Experiment results demonstrate the effectiveness of our proposed framework, providing valuable insights that could significantly advance diagnostic capabilities in Alzheimer's research.

## 5. ACKNOWLEDGMENTS

This work was supported by National Institutes of Health (R01AG075582 and RF1NS128534)

## 6. COMPLIANCE WITH ETHICAL STANDARDS

This is a computational simulation study for which no ethical approval was required.